\newcommand{\beeq}{\begin{eqnarray}}
\newcommand{\eeeq}{\end{eqnarray}}
\newcommand{\be}{\begin{equation}}
\newcommand{\ee}{\end{equation}}

\def\kbo{{\bf k}}

\def\rbo{{\bf r}}
\def\rbop{{\bf r}^\prime}
\def\bbo{{\bf b}}

\def\alfab{{\overline{\alpha}}_s}

\def\gev{\mbox{\rm GeV}}

\documentstyle[11pt,epsfig,a4]{article}

\begin{document}

\titlepage
\begin{flushright}
DESY 01--172 \\ TSL/ISV--2001--0254\\
October 2001
\end{flushright}

\vspace*{1in}
\begin{center}
{\Large \bf Diffusion into infra-red and unitarization  \\ of the BFKL pomeron}\\
\vspace*{0.4in}
K. \ Golec--Biernat$^{(a,d)}$,
L. \ Motyka$^{(b,e)}$
and A.\ M. \ Sta\'sto$^{(c,d)}$ \\
\vspace*{0.5cm}
{\it $^{(a)}$II.\ Institut f\"ur Theoretische Physik,
    Universit\"at Hamburg, Germany}\\
\vskip 2mm
$^{(b)}$ {\it High Energy Physics, Uppsala University, Box 535, 751-21 Uppsala, Sweden} \\
\vskip 2mm
$^{(c)}$ {\it  INFN Sezione di Firenze, Via G. Sansone 1, 50019 Sesto Fiorentino (FI), Italy} \\
\vskip 2mm
$^{(d)}${\it H. Niewodnicza\'nski Institute of Nuclear Physics, Radzikowskiego 152,
 Krak\'ow, Poland} \\
\vskip 2mm
$^{(e)}${\it Institute of Physics, Jagiellonian University, Reymonta 4,
Krak\'ow, Poland} \\
\vskip 2mm
\end{center}
\vspace*{1cm}

\vskip1cm
\begin{abstract}

The BFKL pomeron in perturbative QCD is plagued by the lack of unitarity and
diffusion into  the infra-red region of gluon virtualities.
These two problems are  intimately related.
We perform numerical studies of the evolution equation
proposed by Balitsky and Kovchegov which unitarizes the BFKL pomeron.
We show how diffusion into the infra-red region is suppressed
due to the emergence of a saturation scale and scaling behaviour.
We study universality of this phenomenon as well as its dependence on
subleading corrections to the   BFKL pomeron such as
the running coupling and kinematic constraint. These corrections are very
important for phenomenological applications.

\end{abstract}

\newpage

\section{Introduction}
\label{sec:1}

The Balitsky-Fadin-Kuraev-Lipatov
(BFKL) equation \cite{BFKL} describes the behaviour of  scattering
amplitudes in the limit of high center-of-mass energy $\sqrt{s}$,
in the kinematic range where
perturbative  QCD is applicable.
In  the leading logarithmic approximation (LLA), $\alpha_s \ll 1$ and $\alpha_s \log s \sim 1$,
the behaviour is determined by the leading singularity in the complex
angular  momentum plane at $j=1+4 N_c \alpha_s \ln 2/\pi$,
which corresponds to the vacuum quantum number exchange. In the context of QCD such
exchange is  called the BFKL pomeron.
For $s\rightarrow \infty$, the corresponding total cross section
rises with energy, $\sigma^{tot} \sim s^{\/j-1}=s^{\/0.5}$ for
$\alpha_s\approx 0.2$, violating
the Froissart  bound reflecting
the fundamental principle of unitarity of the
$S$-matrix: $\sigma^{tot}\le C\log(s)$.
The next-to-leading corrections  to the BFKL equation \cite{BFKLNEXT}
reduce the rise of the total cross section but the power-like form in $s$ still holds.

Another cumbersome property of the BFKL equation is the diffusion
of gluon virtualities (dominated by the transverse momentum) into
the infra-red region. The BFKL pomeron is made up of two interacting
reggeized gluons, diagrammatically represented by a set of gluon ladders.
With increasing energy, the distribution of the gluon virtualities
along the ladder quickly enters the low momentum region around
$\Lambda_{\rm QCD}$ with the characteristic diffusion pattern.
Thus the description  which is rooted in perturbative QCD is not
consistent in the high energy limit.

The intuitive physical picture of new phenomena,
called screening or saturation effects, which are in agreement with the unitarity
bound at high energy was provided in \cite{GLR}.
In the context of the BFKL diffusion  the picture looks as follows.
The transverse size of  gluons with transverse momentum $k_\perp$
is proportional to $1/k_\perp^2$.  For large $k_ \perp$, the BFKL mechanism
of gluon radiation, $g\rightarrow gg$, populates the transverse space with
large number of small size gluons per unit of rapidity $Y\sim \ln s$.
The same mechanism also applies to large size gluons with small transverse momenta.
In this case, however, the BFKL approach is incomplete since the gluons strongly
overlap and fusion processes, $gg\rightarrow g$, are equally important.
After taking these processes into account, unitarity is restored by
taming
the rise of the gluon density with transverse momenta  below some
energy-dependent scale $Q_s(Y)$, called saturation scale.
The emergence of such  scale is a fundamental property of the
parton saturation phenomenon.
The saturation scale rises monotonically with $s$,  and for large enough
rapidity $Y$ for which $Q_s(Y)\gg \Lambda_{QCD}$, the presented approach to the high energy
scattering remains consistent. In this sense the solution to the unitarity problem cures
in turn the infra-red diffusion problem.

The  parton saturation idea is realized with the help of non-linear evolution
equations
in which the gluon splitting is described by a linear term while a negative
non-linear term results from
the competing gluon recombination.  The first equation of this type for the
gluon density  was postulated in \cite{GLR} in the
double logarithmic approximation.
More rigorous derivation was given in \cite{MUELLERQIU}.
The attempts to go beyond the double-log asymptotics resulted in
formulations based on semi-classical QCD methods \cite{MCLERVEN,JAL}
since the region of parton saturation is characterized by large
values of colour fields.
A different approach to unitary generalization of the
BFKL equation was proposed by Balitsky \cite{BAL1,BAL2,BAL3}.
By using the operator product expansion
for high energy scattering in QCD, he derived an infinite hierarchy of
coupled equations for $n$-point Wilson-line operators.
Weigert managed to simplify the form of these equations
by writing them as a functional  evolution equation for the
generating functional of the Wilson-line operators \cite{WEIGERT}.
The connection between
the effective theory for the Colour Glass Condensate \cite{MCLERVEN,JAL}
and the evolution equation found by  Weigert has been established in
\cite{CGC1}. Another derivation of the small-$x$ non-linear
equation  has been recently given in \cite{MUELLERNEW}.

The  Balitsky's equations decouple in the large $N_c$ limit. In this limit,
the  equation for the 2-point  function
was independently derived by Kovchegov \cite{KOV1} in the dipole
picture \cite{DIPOL}. The equation unitarizes the BFKL equation by including
a  quadratic term determined by the three-pomeron vertex \cite{BARWUE}. The
properties of this equation were investigated in
\cite{KOV2}-\cite{BRAUN4}
supporting the picture of parton saturation.
The equation introduces the  saturation scale below which
the non-linear effects lead to saturation of the gluon density.
Recently, an extended version of these equations which include
cubic terms has been derived \cite{BAL4}.

In this paper we concentrate on numerical studies of the
Balitsky-Kovchegov evolution equation \cite{KOV1}.
We illustrate the emergence of the saturation
region with the saturation scale and
scaling behaviour of the resulting cross sections.
Such scaling was found in the DIS data at small $x$ \cite{SGK},
after being inspired by the saturation model \cite{GBW1}.
We contrast the  properties of the  non-linear BK equation with those resulting
from the linear BFKL equation. In particular, we show the effect of unitarization on
infra-red diffusion.
These results serve as the starting point for the analysis of   the  Balitsky-Kovchegov
equation
with subleading corrections given
by the running QCD coupling and kinematic constraint.
Although subleading, these corrections
are very important for phenomenological applications
and we carefully examine their impact onto  the universality of the parton
saturation phenomenon.

\section{The Balitsky-Kovchegov equation}
\label{sec:2}

The Balitsky-Kovchegov (BK) equation was derived for
the deep inelastic scattering
of virtual photon on a large nucleus by the resummation of multiple pomeron
exchanges in the leading logarithmic approximation
(when $\alpha_s \ll 1$,
$\alpha_s \ln(1/x)\sim 1$ and
$x\simeq Q^2/s$)  and in  the large $N_c$ limit \cite{KOV1,KOV2}.

The physical picture of this process is provided in the rest frame of the nucleus.
In the small $x$ limit, the virtual photon splits into a $q\bar{q}$  pair long
before the interaction with the  nucleus.
Then the $q\bar{q}$ pair interacts with the nucleus.
In this case the nucleus structure function is given by
\begin{equation}
F_2(x,Q^2)\,=\,
\frac{Q^2}{4\pi^2 \alpha_{em}}
\int d^{\/2}\rbo\, dz
\left\{
|\Psi_T(z,\rbo,Q^2)|^2+|\Psi_L(z,\rbo,Q^2)|^2\right\}\, \hat\sigma (x,\rbo)
\label{eq:f2}
\end{equation}
where  $\Psi_{T,L}$ are the light-cone wave functions for the transversely
and longitudinally polarized virtual photon, and $\hat\sigma$  is the dipole cross
section describing the interaction of the $q\bar{q}$ dipole with the nucleus.
Here,  $\rbo=\rbo_1-\rbo_0$ is the transverse size of the dipole
($\rbo_1,\rbo_0$  are the transverse coordinates of the quark and  antiquark) and
$z$ is  the longitudinal fraction of the photon's momentum carried by the quark.
The sum of the squares of the photon wave functions (for massless quarks) is given by
\begin{equation}
|\Psi_T|^2+|\Psi_L|^2=
\frac{N_c\/ \alpha_{em}}{2\pi^2}\,\sum_f e_f^2
\left\{
[z^2+(1-z)^2]\,
\epsilon^2\, K_1^2(\epsilon r)
+
4\, Q^2 z^2 (1-z)^2\, K_0^2(\epsilon r)
\right\}
\label{eq:wave}
\end{equation}
where  $\epsilon^2=z (1-z)\,Q^2$, $K_{0,1}$ are the Bessel--Mc Donald functions and
the sum is performed over all active quark
flavours $f$. The dipole cross section is related to
the imaginary part of the forward scattering
amplitude of the $q\bar{q}$ dipole on the nucleus
$N(\rbo,\bbo,Y)$,
\begin{equation}
\hat\sigma (x,\rbo)\,=\,2 \! \int d^{\/2}{\bbo }\,N(\rbo,\bbo,Y),
\label{eq:dipole}
\end{equation}
where $\bbo=(\rbo_1+\rbo_0)/2$ is the impact parameter of   the $q\bar{q}$  dipole.

The BK equation is an evolution equation for the amplitude  $N(\rbo,\bbo,Y)$.
In the derivation  \cite{KOV1} it was assumed that
the $q\bar{q}$ pair develops a cascade of gluons which, in the
large $N_c$ limit, form a system of $3$-$\bar{3}$ dipoles. Each dipole
interacts independently with
nucleons in the nucleus via two-gluon exchange.
In this way the following non-linear evolution equation was found
\beeq
\nonumber
\frac{\partial N(\rbo,\bbo,Y)}{\partial Y}
&=&
\overline{\alpha}_s\
(K\otimes N)(\rbo,\bbo,Y)
\\
&-&
\overline{\alpha}_s\,
\int \frac{d^2\rbop}{2\pi}\,
\frac{r^2}{r^{\/\prime 2}(\rbo+\rbop)^2}\,
N(\rbo+\rbop,\bbo+\frac{\rbop}{2},Y)\;N(\rbop,\bbo+\frac{\rbo+\rbop}{2},Y).
\label{eq:kov}
\eeeq
where $\overline{\alpha}_s=N_c \alpha_s/\pi$, and the linear term is
determined by the BFKL kernel
\begin{equation}
\\
\label{eq:bfklkernel}
(K\otimes N)(\rbo,\bbo,Y)=
\int \frac{d^2\rbop}{\pi\/ r^{\/\prime 2}}\,
\left\{
\frac{r^2}{(\rbo+\rbop)^2}\,N(\rbo+\rbop,\bbo,Y)
-
\frac{r^2}{r^{\/\prime 2}+(\rbo+\rbop)^2}\,N(\rbo,\bbo,Y)
\right\}.
\\
\end{equation}
Thus in the linear approximation, when each dipole scatters
only once off
the nucleus, the BFKL equation in the dipole picture is obtained.
The non-linear term in (\ref{eq:kov}) takes into account multiple scatterings
and is essentially determined by the triple pomeron vertex \cite{BARWUE}
in the large $N_c$ limit.
Eq.~(\ref{eq:kov}) unitarizes the BFKL pomeron in the sense that at
$x\rightarrow 0$ and $Q^2$ fixed,
\begin{equation}
F_2\, \sim\, Q^2\, \ln(1/x)\,.
\end{equation}
Thus the power-like
rise with energy for the BFKL pomeron is tamed \cite{KOV2}.

In the infinite momentum frame, the BK equation resums
fan diagrams with the BFKL ladders, corresponding to the pomeron,
splitting into two ladders. This type of summation was originally
proposed by Gribov, Levin and Ryskin (GLR) \cite{GLR} in the double
logarithmic approximation (DLA):
$\alpha_s \ll 1$
and $\alpha_s \ln(1/x) \ln(Q^2/\Lambda_{QCD}^2) \sim 1$.
It was  shown in \cite{KOV1} that eq.~(\ref{eq:kov})
reduces to the GLR equation for the integrated gluon distribution.

The BK equation was derived for a large nucleus by the
summation of the leading contributions in powers of  the atomic number
of the nucleus $A$. Other contributions, e.g. those corresponding
to loops built out of the BFKL ladders, are suppressed by
powers of $A$.
However, this equation can be applied to a nucleon
as a model which hints towards the most important
contributions to the structure function
at small $x$.
The success of the saturation model \cite{GBW1} in the description of the
small-$x$ DIS data
is an indication that this equation might also properly
identify the leading contribution for a nucleon.
\begin{figure}[t]
  \vspace*{-2.0cm}
     \centerline{
         \epsfig{figure=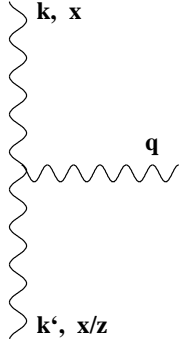,width=9cm}
           }
\vspace*{-2.0cm}
\caption{\it
Kinematics of the gluons in the BFKL ladder.
$x$ and $x/z$ are the longitudinal momentum fractions, and
$k$, $k^\prime$ and $q=k-k^\prime$ are transverse momenta of the gluons.
\label{fig:1}}
\end{figure}

A substantial complication of the analysis of eq.~(\ref{eq:kov}) is
related to the coupling between the $\rbo$ and $\bbo$ variables.
In our presentation
we follow the approximation made in  the earlier analyses that
the dominant contribution to $F_2$ is given by the tranverse sizes $r$
for which
$r\ll b < R$, where $R$ is the nucleus size. In this case  the coupling
between $r$ and $b$ can be neglected and eq.~(\ref{eq:kov}) becomes
independent of $b$. Thus,
the $b$-dependence can be suppressed in $N(\rbo,\bbo,Y)$.
It can then   enter only through the initial condition for the evolution
equation. In a  more refined analysis, however, the exact $b$-dependence
should be taken into account.

For an
azimuthally symmetric solution, $N(\rbo,Y)=N(r,Y)$,  eq.~(\ref{eq:kov}) can be
rewritten  in momentum space in a much simpler form after
performing the Fourier transform \cite{KOV2}
\begin{equation}
\phi(k,Y)
\,=\,
\int \frac{d^2\rbo}{2\pi} \exp(-i\kbo\cdot \rbo)\,
\frac{N(r,Y)}{r^2}
\,=\,
\int_0^\infty
\frac{dr}{r}\, J_0(k\/r)\,N(r,Y),
\end{equation}
where $J_0$ is the Bessel function.
In this case the following equation
is obtained
\beeq
\label{eq:newkov}
\frac{\partial \phi(k,Y)}{\partial Y}
\,=\,
\overline{\alpha}_s\, (K^\prime\otimes \phi)(k,Y)
\,-\
\overline{\alpha}_s\, \phi^{\,2}(k,Y),
\eeeq
and the action of the BFKL kernel (appropriately
shifted in the space of the Mellin moments in $k^2$) is given by
\begin{equation}
(K^\prime\otimes \phi)(k,Y)
\,=\,
\int_0^{\infty} \frac{dk^{\prime 2}}{k^{\prime 2}}\,
\left\{
\frac{k^{\prime 2}\,\phi(k^{\prime},Y)\, -\, k^2\, \phi(k,Y)}
{|k^2\,-\, k^{\prime 2}|}
\,+\,
\frac{k^2\, \phi(k,Y)}{\sqrt{4 k^{\prime 4}\,+\,k^4}}
\right\} ,
\end{equation}
where now $k$ and $k^{\prime}$ are the transverse momenta of the exchanged gluons
in the BFKL ladder, see Fig.~\ref{fig:1}.
The form  (\ref{eq:newkov})
of the BK equation serves as the starting point of our analysis.

\section{Saturation scale and scaling}
\label{sec:3}

In this section we present some interesting properties of the solutions
of the Balitsky-Kovchegov equation, based on a numerical analysis of
eq.~(\ref{eq:newkov}).
In particular, we illustrate how the BFKL diffusion  is suppressed due to
non-linear  effects leading to emergence of a saturation scale.
This issue was also studied in
Refs.~\cite{BRAUN1}-\cite{LUB}.

When studying diffusion in transverse momenta, it is  convenient to
illustrate the results in terms of the function  $\phi(k,Y)$
multiplied by $k$. The reason is very simple. In the linear case,
when the non-linear term in eq.~(\ref{eq:newkov}) is neglected,
the resulting BFKL equation\footnote{
The kernel $K^\prime$ in eq.~(\ref{eq:newkov})
has a saddle point at $\gamma=-1/2$ in the Mellin
space, thus $\phi \sim (k^2)^{-1/2}$.} has the asymptotic  solution which
is proportional to $k^{-1}$. The remaining term
exhibits
the well know diffusion pattern   in the variable $\eta=\ln(k/k_0)$. Thus  for
\begin{equation}
\label{eq:bfklas}
k\,\phi(k,Y) \,=\,
\exp(\alfab \chi(0) Y)\,
\frac{1}{\sqrt{\pi \alfab \chi^{\prime\prime}(0) Y}}\,
\exp\left(
-\,\frac{\ln^2(k^2/k_0^2)}{2 \alfab \chi^{\prime\prime}(0) Y}
\right),
\end{equation}
where $\chi(0)=4\ln 2$ and   $\chi^{\prime\prime}(0)=28\,\zeta(3)$,
diffusion is clearly  isolated. We will contrast the behaviour (\ref{eq:bfklas})
with the solution of the non-linear BK equation.

In order to solve eq.~(\ref{eq:newkov})  an  initial condition
at $Y=0$ has to be specified.
We start from the condition which is concentrated
at some value $\eta_1$ in the logarithmic variable $\eta$
\begin{equation}
\label{eq:start1}
k\,\phi(k,Y=0)\,=\,\delta(\eta-\eta_1).
\end{equation}
In Fig.~\ref{fig:2} we show the solutions
for different values of $Y$ for the linear (BFKL) and the non-linear (BK) equation.
The BFKL solutions illustrate the features of eq.~(\ref{eq:bfklas}),
i.e the  unlimited rise with rapidity (energy)
and diffusion into  the domain of
large and small values of the transverse momentum $k$.
For the solutions of the
BK equation, however,
the small $k$ domain is strongly suppressed and
the rise with energy is significantly
bounded. For large values of $k$, when the non-linearity in the BK equation is small,
the BK and BFKL solutions  coincide.

The impact of unitarization of the BFKL pomeron on the infra-red behaviour
is also illustrated  in Fig.~\ref{fig:3}. These three dimensional plots
show the function
\begin{equation}
\label{eq:norm}
\Psi(k,Y)\,=\,
\frac{k\,\phi(k,Y)}
{k_{max}(Y)\,\phi(k_{max}(Y),Y)},
\end{equation}
where $k_{max}(Y)$ is the value of the
transverse momentum at which $k\,\phi(k,Y)$ has maximum\footnote{
Notice that $\Psi(k,Y)$ is no longer a solution of the non-linear eq.~(\ref{eq:newkov}).}.
Thus,
$k\,\phi(k,Y)$ is re-normalized to unity at $k=k_{max}$.
For the linear BFKL equation, when the asymptotic form
sets in, $\Psi(k,Y)$ is given by the second exponent in  (\ref{eq:bfklas}).
In this case $k_{max}$ is independent of $Y$ and equal to the scale
introduced by the initial condition,
\begin{equation}
k_{max}(Y)\,=\,k_1.
\end{equation}
For the non-linear case, $\Psi(k,Y)$ moves
away of the infra-red region with  $k_{max}(Y)$ depending on $Y$.
This effect is also shown in  Fig.~\ref{fig:4} by making projection
of constant values of   $\Psi(k,Y)$ on the $(\eta,Y)$-plane.
Again,
for small $Y$, when the non-linearity in the BK equation is negligible,
the re-normalized solutions of the BFKL and the BK equations  coincide.
With  increasing Y,  when the
non-linear effects become important, the difference between them
in the region of small $k$  becomes fully visible.

Notice the  parallel straight lines of constant values
of the non-linear solution in a certain region of the plane in   Fig.~\ref{fig:4}.
This means that $\Psi(k,Y)$ in this region is a function of the
combination
\begin{equation}
\label{eq:scalvar}
\xi\,=\,
\ln(k/k_0)\,-\,\lambda\,Y
\,=\, \ln\left(\frac{k}{k_0 \exp(\lambda Y)} \right)~~~~~~~~~~~~~\lambda>0
\end{equation}
(or $e^{\,\xi}$) instead of $k$  and $Y$ separately.

The parameter $\xi$ enumerates different
straight lines  and it is a matter of convention which line corresponds to
$\xi=0$. By a simple rescaling $\xi\rightarrow \xi+\xi_0$,
(equivalent to the transformation $k_0\rightarrow k_0 e^{-\xi_0}$)
we can achieve that
the straight line part of $k_{max}(Y)$ corresponds to $\xi=0$.
In this case  the re-normalized solution is a function of only one variable
$k/k_{max}(Y)$.
The solution of eq.~(\ref{eq:newkov})
has the same scaling property,
\begin{equation}
\phi(k,Y)=\phi(k/k_{max}(Y)) \; ,
\label{eq:kmaxscal}
\end{equation}
if and only if
\begin{equation}
\label{eq:crticond}
\phi(k_{max}(Y),Y)\,=\, \mbox{\rm const} \; ,
\end{equation}
as a function of $Y$.
As shown  in Fig.~\ref{fig:5},
this condition holds true for sufficiently large values of $Y$.
In this case the saturation scale can be defined
\begin{equation}
\label{eq:satscale}
Q_s(Y)\equiv k_{max}(Y)\,=\,Q_0 \exp(\lambda Y)\,,
\end{equation}
with the exponential
dependence on rapidity governed by the value of the scaling parameter
$\lambda$.
The pre-factor
$Q_0$ depends on the initial condition for
eq.~(\ref{eq:newkov}) while the  $Y$-dependence is universal. This is
illustrated in Fig.~\ref{fig:6} where
the slope of the straight lines (i.e. the value of $\lambda$)
in the scaling region is the same for two different
initial distributions,  i.e. that given by eq.~(\ref{eq:start1}) and the
Gaussian distribution
\begin{equation}
\label{eq:start2}
k\,\phi(k,Y=0)
\,=\,
\exp\left[-(\eta-\eta_{1})^2/\beta^2
\right].
\end{equation}
We have verified this statement for different values
of the parameters $\eta_{1}$ and $\beta$.

It is tempting to claim that for $\xi<0$, i.e. for $k<Q_s(Y)$,
scaling holds while for  $\xi>0$ it is violated.
The sharp boundary, however, does not appear in reality and the transition
between these two regions is smooth. This is shown in Fig.~\ref{fig:7} where
the function $k/Q_s(Y)\,\phi(k,Y)$ is plotted as a function of the scaling
variable $k/Q_s(Y)$ for different values of the rapidity $Y$.
The scaling behaviour is
represented by a common line to the left of the point $k/Q_s(Y)=1$. The slow departure
from scaling for $k>Q_s(Y)$ is clearly visible.
In this sense the line $k=Q_s(Y)$ is
only an approximation characterizing the position of the transition region
in the $(k,Y)$-plane. However, the choice based on the position in $k$ of the
maximum of $k\phi(k,Y)$ as a function of $Y$ is the most natural one.

Let us finish by recalling that the scaling property of the solution  of
the BK equation results in the scaling law for
the $\gamma^*p$ total cross section  $\sigma_{\gamma^*p}\sim F_2/Q^2$,
\begin{equation}
\sigma_{\gamma^*p}(x,Q^2)\,=\,\sigma_{\gamma^*p}(Q^2/Q_s^2(Y)).
\end{equation}
Such  scaling law (``geometric scaling'') was found in
the small-$x$ DIS data \cite{SGK}.
\section{Analytical insight}
\label{sec:3a}
The emergence of scaling and  the  saturation scale
may also be investigated using the analytic
method proposed in \cite{BL}.
Let us introduce the Mellin transform of the solution $\phi(k,Y)$
of eq.~(\ref{eq:newkov}):
\begin{equation}
\tilde\phi(k,\omega)\, =\, \int dY\, e^{-\omega Y } \phi(k,Y)
\label{eq:mellin}
\end{equation}
where $\omega$ is the Mellin variable conjugated to the rapidity $Y$.
We adopt the following ansatz proposed in ref.~\cite{BL}
\begin{equation}
\tilde\phi(k,\omega)\, =\, \tilde\phi(\omega)\, e^{(\gamma(\omega)-1) t},
\label{eq:phisol}
\end{equation}
where $\gamma(\omega)$ is  anomalous dimension and $t=\ln(k^2/k_0^2)$.
By taking the Mellin transform of eq.~(\ref{eq:newkov}), we find
\begin{equation}
\left\{\omega   - \bar{\alpha_s} \chi(\gamma(\omega))\right\}\, \tilde\phi(\omega)\,
e^{(\gamma(\omega)-1) t}\,=\,-
\overline{\alpha}_s \int {d\omega' \over 2\pi i}\; \tilde\phi(\omega-\omega')\,
\tilde\phi(\omega')\,
 e^{(\gamma(\omega-\omega')+\gamma(\omega')-2) t}\,,
\label{eq:melkov}
\end{equation}
where $\chi(\gamma)$ is eigenvalue of the leading BFKL kernel
\begin{equation}
\chi(\gamma) \, = \, 2 \psi(1)\, -\, \psi(\gamma)\, -\, \psi(1-\gamma) \; ,
\label{eq:eigenvalue}
\end{equation}
and $\psi$ is the Euler digamma function.
The right hand side of eq.~(\ref{eq:melkov}) can be evaluated in
the saddle point approximation \cite{BL}, which results in
\begin{equation}
\left\{\omega   - \bar{\alpha_s} \chi(\gamma(\omega))\right\}\, \tilde\phi(\omega)\,
e^{(\gamma(\omega)-1) t}\,=\,-
\overline{\alpha}_s\,  \tilde\phi^{2}(\omega/2)\,  e^{(2 \gamma(\omega/2)-2) t}\,
{1 \over \sqrt{4\pi\gamma''(\omega/2) t}}\,.
\label{eq:kovsaddle}
\end{equation}

In the region where the non-linear term  in (\ref{eq:kovsaddle}) can be neglected,
the anomalous dimension $\gamma(\omega)$
fulfils the equation
\begin{equation}
\label{eq:pert}
\omega   - \bar{\alpha_s} \chi(\gamma(\omega)) \,=\,0\,.
\end{equation}
In the double logarithmic approximation,
the dominant contribution to $\chi(\gamma)$ comes
from the pole $1/\gamma$, and the solution for the anomalous dimension
reads
\begin{equation}
\gamma(\omega) \, = \, {\alpha_s}/{\omega}\,.
\label{eq:soldoublelog}
\end{equation}
On the other hand, in the saturation region, where the nonlinearity
cannot be neglected, the following condition
is obtained from eq.~(\ref{eq:kovsaddle})
\begin{equation}
\gamma(\omega) \, = \, 2 \gamma\left({\omega/2}\right) - 1\,,
\label{eq:anomcond}
\end{equation}
which is satisfied by
\begin{equation}
\gamma(\omega)\,=\, C \, \omega + 1 \; ,
\label{eq:anomsol}
\end{equation}
where $C$ is a  constant \cite{BL}.

Using (\ref{eq:anomsol}) and (\ref{eq:phisol}) and performing
the inverse  Mellin transform  in $\omega$,
we find  the solution to the BK equation in the saturation region
\begin{equation}
\phi(k,Y) \,=\,
\int {\frac{d\omega}{2\pi i}}\, e^{\omega Y}
\tilde\phi(k, \omega)\,
\, = \, \int {\frac{d\omega}{2\pi i}}\, \tilde\phi(\omega)\,
e^{\omega (Ct+Y)}
\label{eq:phisolscal}.
\end{equation}
We see that the solution $\phi(k,Y)$ depends only on the following combination,
\begin{equation}
C\,t+Y\,=\,2 C\, \left[ \ln(k/k_0)+\frac{1}{2C} Y \right],
\label{eq:levinsat}
\end{equation}
which is related in a simple way to the previously defined quantity $\xi$,
eq.~(\ref{eq:scalvar}), with the scaling parameter $\lambda=-1/(2C)$.
In this way the scaling behaviour, seen in the numerical
analysis presented in the previous section, is found.

The solutions to eqs.~(\ref{eq:pert}) and  (\ref{eq:anomcond}),  as well
as their derivatives with respect to $\omega$,
should match at some $\omega=\omega_c$ (see \cite{BL} for a full discussion).
This allows to fix the value of the unknown parameter
$C$ in (\ref{eq:levinsat}).
From (\ref{eq:anomsol}) one finds
\begin{equation}
\frac{d\gamma(\omega)}{d\omega}\,=\,\frac{\gamma(\omega)-1}{\omega},
\end{equation}
and the same relation should be valid for the solution $\gamma(\omega)$
of eq.~(\ref{eq:pert}) at $\omega=\omega_c$. Thus,  by differentiation of
eq.~(\ref{eq:pert}) with respect to $\omega$, we find
\begin{equation}
\label{eq:critline}
\frac{d \chi(\gamma_c)}{d\gamma}\,=\,-\frac{\chi(\gamma_c)}{1-\gamma_c}\,,
~~~~~~~~~~~~~~~~~~~~~~
\omega_c\,=\,\ \overline{\alpha}_s\chi(\gamma_c)\,.
\end{equation}
In the DLA, when $\chi(\gamma)=1/\gamma$,
one obtains
\begin{equation}
\gamma_c  =  {\frac{1}{2}}\,,
~~~~~~~~~~~~~~~~
\omega_c  =  2\, \overline{\alpha}_s\,,
~~~~~~~~~~~~~~~~
C  =  {-\frac{1}{4\, \overline{\alpha}_s}}\,.
\label{eq:critresult}
\end{equation}
Thus the scaling parameter
$\lambda=-1/(2C)=2\,\overline{\alpha}_s$.
In the case
of the BK equation with
the full BFKL kernel,
we found numerically that the scaling parameter equals
\begin{equation}
\lambda \,=\, 2.05 \, \overline{\alpha}_s\,.
\label{eq:avalue}
\end{equation}
This value is very close to the DLA value and the results quoted
in \cite{LEVTUCH2}  and \cite{BRAUN4}.
We have also checked that (\ref{eq:avalue}) still holds
in the case of collinear approximation to the BFKL kernel, i.e. keeping
only $1/\gamma$ and $1/(1-\gamma)$ poles in eq.~(\ref{eq:eigenvalue}).

Summarizing,  the non-linear BK equation has three regimes
depending on the relative importance of the non-linear term:
(I) the linear regime (when $k^2>Q_s^2(Y)$)  is dominated by
the conventional BFKL evolution characterized by the BFKL
intercept and diffusion in transverse momenta.
In the DLA, the equation is driven by the $1/\gamma$ pole of the kernel
eigenvalue (\ref{eq:eigenvalue});
(II)
in the saturated   regime ($k^2<Q_s^2(Y)$), the non-linear effects are strong and the gluon
density is saturated.  In this case, in the linear part of the BK  equation
the $1/(1-\gamma)$ pole of $\chi(\gamma)$ is important;
(III)
the  transition between the linear and
saturated regimes is defined by the matching (\ref{eq:critline})
which  depends on the detailed form of the evolution kernel.

\section{The running coupling effects}
\label{sec:4}


The inclusion of the running QCD coupling (RC) into the leading order BFKL formalism
can be treated as a  partial, phenomenological way to account for
a part of important non-leading effects.
 The results of the next-to-leading logarithmic  calculations
\cite{BFKLNEXT} may be invoked to motivate such approach \cite{CCS1,SALAM,CCS2}.
In the context of the infra-red diffusion, the RC BFKL equation
exposes serious problems,  in particular, large sensitivity to the treatment of
the non-perturbative region. The main result of the forthcoming analysis of the
BK equation with the running coupling
is that in this case   the ambiguities related to the treatment
of the infra-red domain are substantially reduced.

The BFKL equation explicitly takes into account the contribution
from the momenta down to $k^2 =0$. Certainly, this is beyond the
applicability of the perturbative formalism. Moreover, for the
BFKL equation with the running coupling,
a purely mathematical difficulty arises due to the Landau pole
\begin{equation}
\alpha_s(k^2)\, =\, {\frac{4\pi}{b_0 \,\ln{ (k^2 / \Lambda^2)}}}.
\label{eq:running}
\end{equation}
with $b_0=11- 2 n_f/3$.
Therefore, a cut-off $k_0$ for gluon virtualities has to be introduced
below which $\alpha_s$ is frozen or set to vanish.
However, the BFKL amplitude becomes strongly sensitive to the
choice of the cut-off.
In the case of $\alpha_s(k^2) = 0$ for $k^2 < k_0 ^2$,
the  BFKL kernel with running coupling has a discrete, cut-off dependent spectrum.
The largest eigenvalue $\lambda_{max}$, which governs the high
energy behaviour of the scattering amplitude, is bounded by \cite{COLKWIE}:
\begin{equation}
\frac{3.6}{\pi}\alpha_s(k_0 ^2)\, \leq\, \lambda_{max}\, \leq \,
                     {\frac{12 \ln{2}}{\pi}}\alpha_s(k_0 ^2).
\label{eq:CK}
\end{equation}
From this, one concludes that the equation is unstable, i.e. the pomeron
intercept may reach an arbitrarily large value depending on
the infra-red cut-off.
Besides, it was found that the
solution $\phi (k,Y)$ of the BFKL equation with the running coupling
no longer exhibits the diffusion pattern typical  for the fixed
coupling  case. Namely, it takes the following factorized form for sufficiently
large $Y$ and $k^2>k_0^2$ \cite{LUNDRC}:
\begin{equation}
k \phi (k,Y)\,\sim\,
\exp\{\lambda Y\}\, {\frac{1}{k}}\,[\ln \left({k^2}/{k_0^2}\right)]^{\nu},
\label{eq:phi_lin_rc}
\end{equation}
with $\lambda=\lambda_{max}$ and  $\nu = 12/(11\lambda) - 1$, so that
$\nu$ is positive for moderate $\lambda$ and becomes negative
for $\lambda > 12/11$.
Thus, diffusion into infra-red is strongly enhanced in the running coupling case,
keeping the gluon chains in the region of small virtualities.
As a result, the high energy behaviour is governed by the
non-perturbative part of the gluon configurations.
However, it is still possible to perform collinear
factorization in which the genuine non-perturbative contribution may be
absorbed into the input for the evolution in $Q^2$ \cite{KWIECMAR}.

The BK equation is also of perturbative origin so its validity
at low $k^2$ may be questioned. However, the equation is  strongly constrained
by unitarity requirements. In particular, the demand for the dipole
scattering amplitude to saturate to unity for large dipoles
(the scattering matrix in the forward direction $S=0$ for large dipoles)
results in a universal form of $\phi(k,Y) \sim \ln k$ for small $k$ and
fixed $Y$.
The  statement about the large $r$ behaviour of the dipoles
is quite general. Interestingly,
the Balitsky-Kovchegov equation,
which is rooted in  perturbative QCD,
ensures such behaviour of the solutions at large $r$.

This conclusion remains unaltered also in the scenario with the running coupling.
The unitarity effects
are strongest in the region of small values of the gluon momenta.
This is due to the fact that
the  solution to the linear part of eq. (\ref{eq:newkov}) is proportional
to $k ^{-1}$ in the fixed coupling case, (eq.~{\ref{eq:bfklas}), and it
rises faster towards low $k$ making the non-linear  rescattering term
largest for small $k$.
This effect tames the rapid rise of the amplitude governed by
the running coupling taken at the cut-off scale $k_0^2$.
The evolution for low $k$ is
essentially frozen to the known form and it is the perturbative region
of $k$ which drives the evolution. This was  noticed in \cite{BRAUN3}
as a supersaturation phenomenon.

In Fig.~\ref{fig:8}~a,b we show the results of the numerical investigation of
the BK  equation  (\ref{eq:newkov}) with the running coupling.
We considered two scenarios: the cut-off regularization of the running coupling,
\begin{equation}
\alpha_s(k^2)=0~~~~~~~~~~\mbox{\rm for}~~~~~~~~~~ k^2<k_0^2\,,
\label{eq:sc1}
\end{equation}
and   freezing of the  running coupling  at low scales
\begin{equation}
\alpha_s\,(k^2)\,\,\,\rightarrow \,\,\,\alpha_s(k^2+k_0^2)\,.
\label{eq:sc2}
\end{equation}
The latter option is supported by
some solid theoretical arguments \cite{Simonov} and
the QCD lattice data,
see e.g. ref.~\cite{Badalian}. The argument $k^2$ is given by the virtuality
of the outgoing exchanged gluon, see Fig.~\ref{fig:1}, and the same
$\alpha_s(k^2)$ is assumed in the linear and non-linear terms.
A different choice of the scale  would correspond to a
 modification of the BFKL kernel in the NLL approximation
\cite{BFKLNEXT,CCS1} which
does not influence significantly the properties of the solution.
In  order to investigate the sensitivity to
infra-red details,
two values of  the cut-off  were used: $k_0^2 = 0.1~\mbox{\rm GeV}^2$~(a)
and  $k_0^2 = 1~\mbox{\rm GeV}^2$~(b). Both scenarios for the running coupling
lead to similar results.  The contours of constant values of the
function $\Psi(k,Y)$, eq.~(\ref{eq:norm}),
in the scenario (\ref{eq:sc1}) are
plotted in the $(\eta,Y)$-plane.  The input is given by the
Gaussian (\ref{eq:start2}) centered at $k_1=10^2$~GeV.

We see in Figs.~\ref{fig:8} that the solution to  the BFKL equation
passes through a rapid ``tunnelling'' transition
(the detailed discussion of this phenomenon is presented in \cite{CCS2}),
collapsing from the region of $k \simeq k_1$(input scale)  to $k \simeq k_0$
(cut-off scale) at certain rapidity.
As expected, for larger $Y$ the shape in $k$ factors out from the
$Y$-dependence for the solution of the BFKL equation $\phi(k,Y)$.
In the  re-normalized solution $\Psi(k,Y)$, eq.~(\ref{eq:norm}),
the $Y$-dependence drops out and the lines of constant
values become parallel to the $Y$-axis, as seen in Fig.~\ref{fig:8}.

The behaviour of the solution of the non-linear  BK equation is different.
Now, we distinguish three regimes for $\Psi(k,Y)$:
\begin{itemize}
\item[I.]  For small Y, the linear  BFKL regime  occurs
with diffusion weighted towards low $k$.
\item[II.] When $Y$ increases, the distributions broadens and
the tunnelling transition occurs.
However, this effects is different from the collapse
observed in the linear BFKL case since the saturation scale is generated
which suppresses  $\Psi(k,Y)$ for small $k$.
\item[III.] For high $Y$, geometric scaling (\ref{eq:kmaxscal}) appears.
 The saturation scale increases with $Y$ and the dynamics of
evolution is governed by perturbative values of $k$.
\end{itemize}

The dependence on the cut-off $k_0$  is marginal for large $Y$ in  both
the saturated and the non-saturated range of $k$.
The statement about the independence from the infra-red parameter $k_0$
also holds for the effective intercept which governs the increase of the gluon
density with $Y$ for the scales much larger than the saturation scale $Q_s(Y)$.

It is interesting to estimate how the running coupling
affects the $Y$ dependence of the saturation scale $Q_s(Y)$.
We adopt a natural approximation that
the local exponent $\lambda(Y)= d \ln(Q_s(Y)/\Lambda)\; /dY$
takes the form $\lambda(Y) = 2 \bar\alpha_s (Q^2_s(Y))$ where $\Lambda=\Lambda_{QCD}$.
The above form is motivated by the leading logarithmic result
with the fixed coupling (\ref{eq:critresult}) and its numerical
verification (\ref{eq:avalue}).
Thus, we have
\begin{equation}
{d\ln (Q_s(Y)/\Lambda) \over dY} = {12 \over b_0 \ln ( Q_s(Y) /\Lambda)}\,,
\end{equation}
%
with the initial condition $Q_s(Y_0) = Q_0$  and  $Y_0$ chosen in the region where
scaling sets in. The solution takes the form
%
\begin{equation}
Q_s(Y) = \Lambda\; \exp\left( \sqrt{{24\over b_0}\, (Y-Y_0) + L_0 ^2} \right)\,,
\qquad Y>Y_0\,,
\label{eq:satscalerc}
\end{equation}
where $L_0 = \ln (Q_0/\Lambda)$.
It follows that the local exponent $\lambda(Y)$ decreases with
increasing rapidity, and $\lambda(Y) \sim 1/\sqrt{Y}$  for very large $Y$.
Such dependence is indeed seen in the numerical analysis.

In Fig. \ref{fig:9} we illustrate scaling in the running coupling case by
showing the function $(k/Q_s(Y)) \, \phi(k,Y)$ plotted versus
$k/Q_s(Y)$ for different values of rapidity.  $Q_s(Y)$ is given  by
formula (\ref{eq:satscalerc}) with the initial condition
$Q_s(Y=0) = 2~\mbox{\rm GeV}$.
The overlapping curves at low values of the scaling variable
clearly indicate
that for $k<Q_s(Y)$  scaling  is satisfied to a very good accuracy,
thus justifying our ansatz (\ref{eq:satscalerc}) for the saturation
scale.

Summarizing,
the Balitsky-Kovchegov equation   avoids the self-consistency problems
of the BFKL equation when the running coupling is included.
The sensitivity to the treatment of the infra-red region is much
smaller than in the linear case due to the appearance of the saturation
scale.
In contrast to the  BFKL equation with the running coupling, the large $Y$
asymptotics of the dipole scattering amplitude arising from the BK
equation is governed by gluon virtualities
in the perturbative domain.

\section{The kinematic constraint effects}
The BK equation (\ref{eq:kov}) has been derived in the
leading logarithmic approximation. It has been argued that the NLL corrections
should be smaller that the saturation effects embodied in this equation.
Since  the corrections from the NLL BFKL kernel become important
at rapidities $Y_{NLL} \sim 1/\alpha_s^{5/3}$, they should be parametrically smaller
than the unitarity corrections  which enter at rapidities of the order
$Y_U\sim 1/\alpha_s \ln (1/\alpha_s)$. However, it is well known that the
next-to-leading  corrections \cite{BFKLNEXT} to the BFKL kernel are numerically
important and even make the high energy expansion unstable. Namely,
the pomeron intercept in the NLL order,
\begin{equation}
\omega_{NLL}\; \simeq \; 4 \ln 2 \;  \overline{\alpha}_s\,
( 1- 6.47\, \overline{\alpha}_s )\,,
\label{eq:interceptnll}
\end{equation}
becomes negative already for small value of $\overline{\alpha}_s \simeq 0.16$ \cite{BFKLNEXT}.
These problems have been solved   in  \cite{CCS1} where a
method which stabilizes the high energy expansion by
a proper resummation of collinear and energy-scale-dependent terms
to all orders has been proposed.
As a  result, one obtains  a
renormalization and resummation scheme independent equation.
Similar approaches have also been developed in \cite{THORNE,ABF}.
In the previous section we have studied the NLL corrections due to the
running of the coupling $\alpha_s$.
Here, we  study different subleading  effects,
namely those  coming from the change of the energy scale.

In order to verify the importance of the NLL terms we shall modify
the BK equation in momentum space (\ref{eq:newkov})
by imposing so called kinematic constraint \cite{AGS,KMS} (see fig. \ref{fig:1}),
\begin{equation}
k'^2\, <\, k^2 /z
\label{eq:kc}
\end{equation}
on the real emission term in the linear part of the BK  equation.
The origin of this constraint lies in the fact that in the
multi-Regge kinematics the exchanged gluon virtuality is dominated by
its transverse momentum, $|k'^2| \simeq |k_T^2|$.
After imposing the kinematical constraint one arrives at the following modified
BK equation
\beeq
\phi(k,Y) =   \phi_0(k)
&+&
\overline{\alpha}_s \int_0^Y dy
\left[
\int_0^{\infty} \frac{dk^{\prime 2}}{k^{\prime 2}}\,
\right.
\left\{
\frac{k^{\prime 2}\phi(k^{\prime},y)\, \Theta(Y-y-\ln (k^{\prime 2}/k^2))\, -\, k^2 \phi(k,y)}
{|k^2\,-\, k^{\prime 2}|}
\right.
\nonumber
\\ \nonumber
\\
& + &
\left.
\left.
\frac{k^2\, \phi(k,y)}{\sqrt{4 k^{\prime 4}\,+\,k^4}}
\right\}
\,  - \, \phi^2(k,y)
\right] \, ,
\label{eq:newkovkc}
\eeeq
where $y=\ln(z/x)$ and $Y-y=\ln(1/z)$, see Fig.~\ref{fig:1}.
The function $\phi_0(k)$ is an initial condition at $Y=0$.
Notice that we no longer deal with
a differential equation in rapidity, compare
eq.~(\ref{eq:newkov}).

Condition (\ref{eq:kc}) leads to the following
modification of the leading eigenvalue (\ref{eq:eigenvalue}) of the BFKL kernel
\begin{equation}
\chi(\gamma,\omega) \; = \; 2\psi(1) - \psi(\gamma) - \psi(1-\gamma+\omega)\,,
\label{eq:chiom}
\end{equation}
where $\gamma$ and $\omega$ are the Mellin variables conjugated to
$\ln k^2/\Lambda^2$ and $Y$, respectively.
The indicated shift in the second $\psi$ function  accounts for
the resummation to all orders of subleading terms
resulting from the changes of the energy scale.
The obtained  $\omega$-dependent kernel  ensures
that the collinear limits are always correctly reproduced \cite{CCS1,SALAM2}.
The new Pomeron intercept  $\omega_{NLL}$
stays always positive and is significantly reduced as compared to the
leading logarithmic value $\omega_{LL}=4\ln 2\,\overline{\alpha}_s$
\cite{CCS1,KMS}.
It also depends  non-linearly on  $\bar{\alpha}_s$ which is a consequence
of the $\omega$  dependence of the  eigenvalue (\ref{eq:chiom}).

The solution of the BK equation with  the kinematic constraint
also exhibits the previously discussed scaling property.
Repeating the analysis of Section~\ref{sec:3a} with the BFKL kernel eigenvalue
$(\ref{eq:chiom})$ instead of $(\ref{eq:eigenvalue})$, we arrived at the
scaling condition in the saturation region with the variable (\ref{eq:levinsat}).
Now, the value of the scaling parameter $\lambda$ is different. The matching condition
(\ref{eq:critline}) is replaced by
\begin{equation}
\label{eq:critlinenew}
\frac{\partial \chi}{\partial\gamma_c}
\,=\,
-\frac{\chi}{1-\gamma_c}\,
\left(
1-\overline{\alpha}_s\,\frac{\partial \chi}{\partial \omega_c}
\right)
\,,
~~~~~~~~~~~~~~~
\omega_c\,=\,\ \overline{\alpha}_s\chi(\gamma_c,\omega_c)\,.
\end{equation}
Let us consider the DL approximation in which
the eigenvalue (\ref{eq:chiom}) is expanded
in $\gamma$ and then in $\omega$ around $\gamma=\omega = 0$
\begin{equation}
\chi(\gamma,\omega) \simeq   {1 \over \gamma} - {\pi^2 \over 6}\omega\,.
\label{eq:chikcl}
\end{equation}
Using this form of the BFKL eigenvalue in the matching conditions (\ref{eq:critlinenew}),
we find
\begin{equation}
\gamma_c  =  {\frac{1}{2}}\,,
~~~~~~~~~~~~~~~~
\omega_c  =  {2\, \overline{\alpha}_s \over 1 + \pi^2\bar\alpha_s /6}\, ,
~~~~~~~~~~~~~~~~
C  =  -{1+\pi^2\overline\alpha_s /6 \over 4\, \overline{\alpha}_s}\,,
\label{eq:critresultkc}
\end{equation}
which should be compared to the result (\ref{eq:critresult}) for the BK equation with
the linear kernel without the kinematic constraint. Now, the scaling parameter
$\lambda=-1/(2C)$ reads
\begin{equation}
\lambda={2\,\overline{\alpha}_s \over 1 + \pi^2\overline\alpha_s /6}.
\label{eq:lambdakc}
\end{equation}
This approximate result suggests that  the
scaling parameter is reduced in comparison with the standard BK equation,
see eq.~(\ref{eq:avalue}).
Moreover, the dependence of $\lambda$ on $\bar\alpha_s$ is non-linear.

We have solved numerically the BK equation (\ref{eq:newkov}) with  the kinematic
constraint in the BFKL kernel
and found that indeed the scaling property approximately  holds.
The scaling parameter $\lambda$ is found to be reduced with respect to
the value (\ref{eq:avalue}) for the BK equation without the kinematic
constraint.
The non-linear dependence on $\bar\alpha_s$ of the form
(\ref{eq:lambdakc}) is confirmed by the numerical considerations
within some $15 \%$ relative deviation, towards smaller values of $\lambda$.

In Fig.~\ref{fig:9} we show the solution $\phi(k,Y)$ of the BK equation with and without
the kinematic constraint as a function of $k$ for different values of $Y$.
In both cases  the same Gaussian input (\ref{eq:start2})
concentrated at $k_1 = 1 \, {\rm GeV}$ is used.
The overall pattern is similar in both cases, with power-like fall for the
large values of $k$ and the $\ln(k/k_0)$ behaviour for small values of $k$.
However, the transition between the two regimes at the same values of rapidity
occurs for smaller values of $k$ in the  case of solution with the
kinematic constraint.
This means that the saturation scale generated by the BK
equation is smaller when the kinematic constraint is present.
This is a phenomenon analogous to the well known reduction of the leading order
BFKL pomeron intercept by including the next-to-leading corrections.

From  Fig.~\ref{fig:9} it is seen that the kinematic constraint is potentially
very important for the phenomenological studies  based on  the
BK equation. Its effect manifests for instance
in a slower increase of the saturation scale than in  the standard case.
The impact of the kinematic constraint is even more important in the linear regime,
where the solution $\phi(k,Y)$
drops by two orders of magnitude at $Y=20$ (for $\alpha_s = 0.2$) when the constraint
is included. However, the general picture of the saturation process remains unaltered.
So, future reliable phenomenological applications of the BK equation to should involve the
kinematic constraint or another representation of resummed non-leading corrections
to the BFKL part of the evolution kernel. It might also be valuable to combine
the kinematic constraint and the running coupling effects in the BK equation.

\section{Summary}

We presented detailed numerical
analysis of the non-linear Balitsky-Kovchegov (BK) equation.
We showed how diffusion into the infra-red domain
for the  linear BFKL equation is suppressed for the BK equation
due to the emergence of the saturation scale $Q_s(Y)$,
which depends on the rapidity $Y$.
The saturation scale divides the $(k,Y)$-plane into a region governed by the linear BFKL
evolution when $k> Q_s(Y)$, and the region in which gluon saturation occurs when
$k< Q_s(Y)$. In the latter region,  the solution is a function of only
one variable $k/Q_s(Y)$, instead of $k$ and $Y$ separately (scaling behaviour).
Our results are fully compatible with
the previous analyses performed at the leading order. The scaling parameter
$\lambda\approx 2\bar{\alpha_s}$ which governs the $Y$ dependence of the saturation scale,
$Q_s\sim \exp{(\lambda Y)}$,
is universal and does not depend on the form of the initial conditions
for the evolution.

We also investigated the influence of the  NLL corrections to the
BFKL kernel such as the running coupling
and kinematic constraint. Similarly to the fixed coupling case, when $\alpha_s$ runs,
the diffusion of gluon virtualities into
the infra-red domain is avoided  by generation of the saturation scale $Q_s$. We found,
however, a different dependence of $Q_s$ on the rapidity, see eq.~(\ref{eq:satscalerc}).
Thus,
the non-linear evolution at high rapidity is driven by
gluon momenta in the perturbative range independently of the
initial condition for the evolution.
The details of the BK solutions in the running coupling case are
quite insensitive to  the   regularization  of   $\alpha_s$ at small scales.

We also studied the impact of the kinematic constraint onto
the non-linear BK equation and found that an approximate scaling property
still holds. However, the kinematic constraint slows down the rise
of the gluon density with increasing rapidity (energy). Thus, the saturation scale
increases slower with rapidity than in the
unconstrained case, see eq.~(\ref{eq:lambdakc}).

We derived and tested numerically the
analytic formulae  describing the dependence of the saturation scale
on rapidity when either the kinematic constraint or the running
QCD coupling are included in the BK equation.
This is the first systematic study of impact of those subleading
corrections on the gluon density saturation  in the
Balitsky-Kovchegov framework. These results suggest that
the subleading effects are very important for the construction of a
reliable phenomenology.

\section*{Acknowledgments}

We would like to thank
Jochen Bartels, Jan Kwieci\'nski, Genya Levin  and Gavin Salam for interesting discussions.
KGB and LM are grateful to  Deutsche Forschungsgemeinschaft and
the Swedish Natural Science Research Council, respectively,
for fellowships.
This research was supported by the EU Framework TMR
programme, contract FMRX-CT98-0194, by the Polish Committee for Scientific Research
grants Nos.\ KBN~2P03B~120~19, 2P03B~051~19, 5P03B~144~20.


\newpage
\begin{figure}[t]
  \vspace*{0.0cm}
     \centerline{
         \epsfig{figure=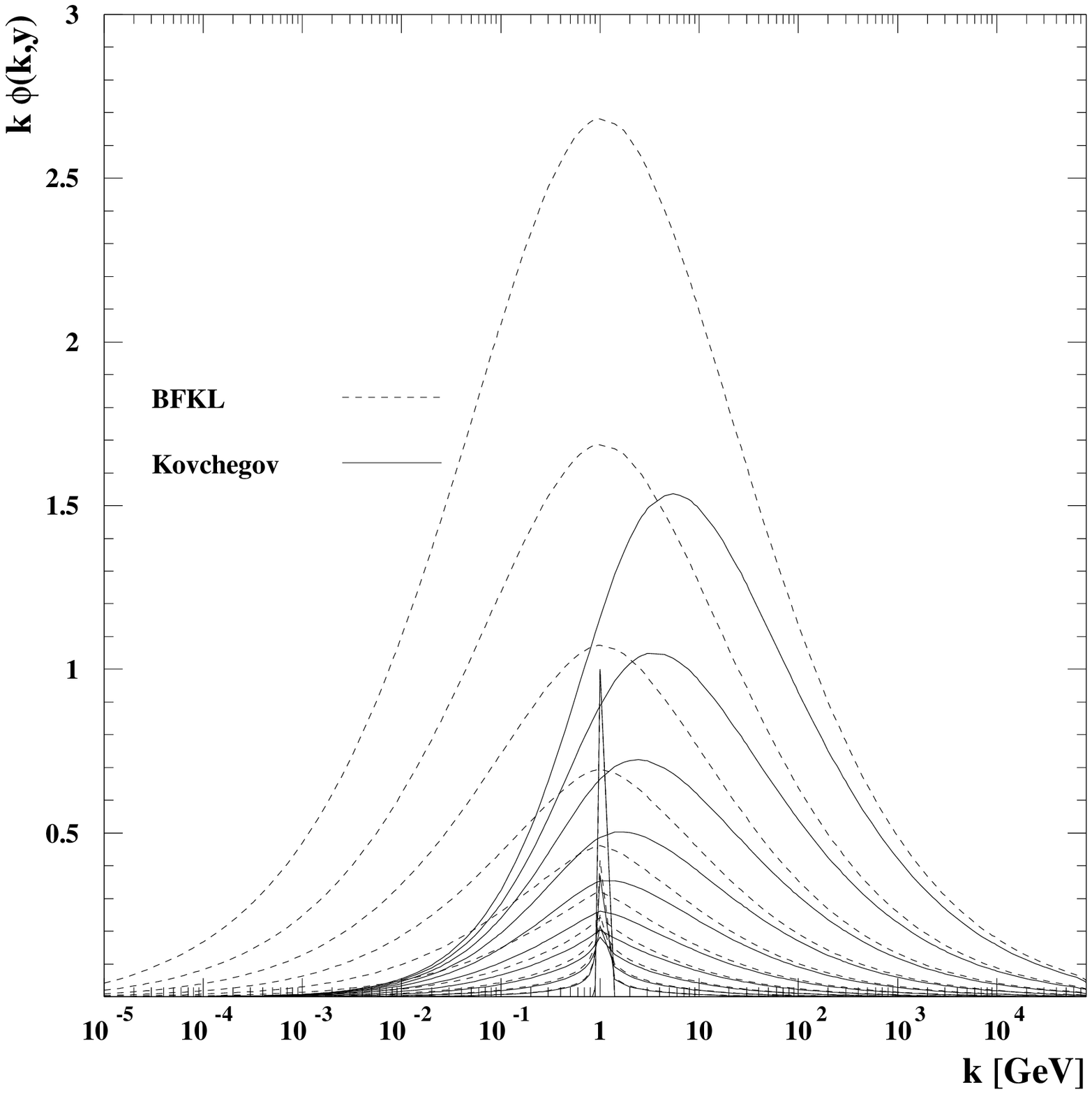,width=15cm}
           }
\vspace*{0.5cm}
\caption{\it
The functions $k\phi(k,Y)$ constructed from solutions to the BFKL and
the Balitsky-Kovchegov equations with the input (\ref{eq:start1})
for different values of the evolution parameter $Y=\ln(1/x)$ ranging from $1$ to $10$.
The coupling constant $\alpha_s=0.2$.
\label{fig:2}}
\end{figure}

\newpage
\begin{figure}[t]
  \vspace*{0.0cm}
     \centerline{
         \epsfig{figure=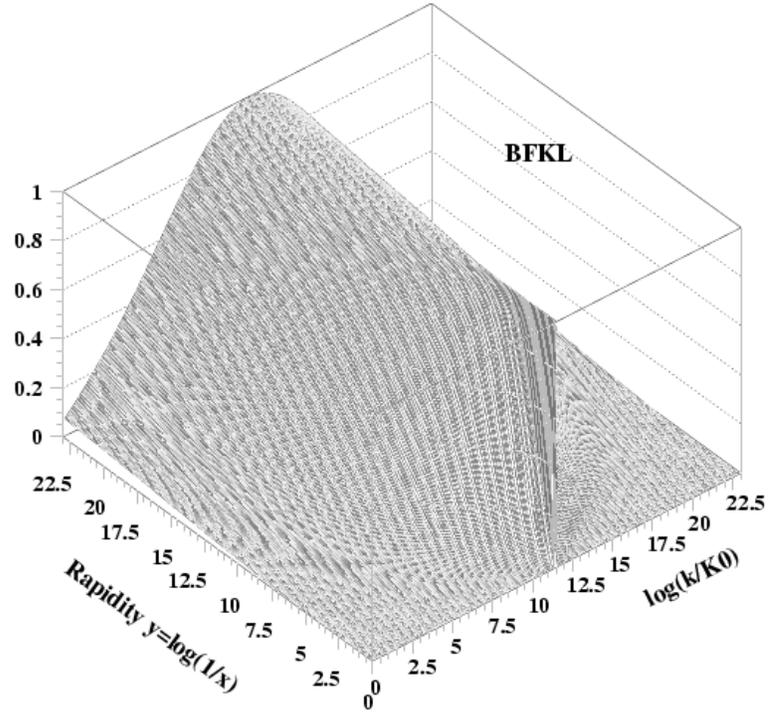,width=10cm}
           }
\vspace*{1.0cm}
     \centerline{
         \epsfig{figure=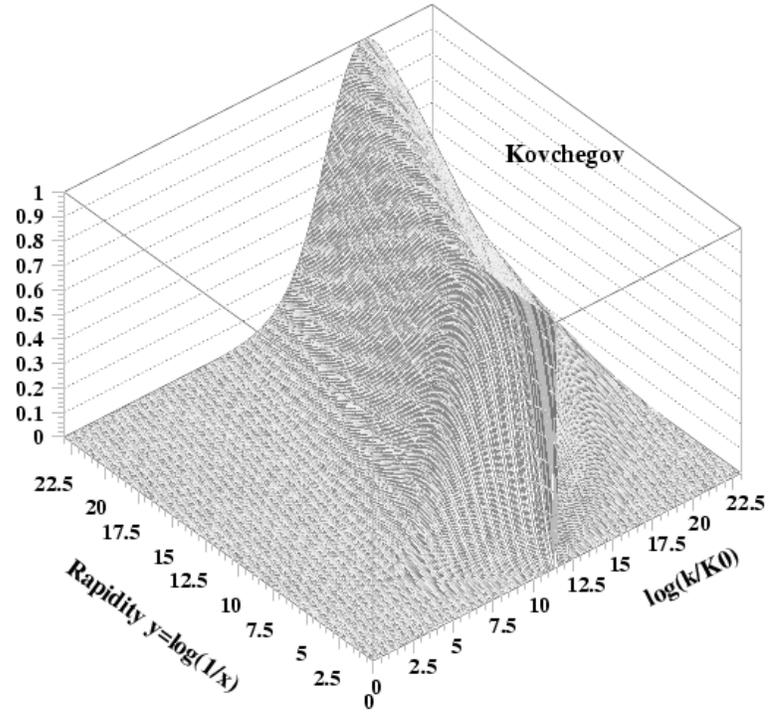,width=10cm}
           }
\vspace*{0.0cm}
\caption{\it The re-normalized solutions $\Psi(k,Y)$ from the input (\ref{eq:start1})
for the BFKL and the Balitsky-Kovchegov equations as a function of the rapidity $Y$ and
$\eta=\ln(k/k_0)$ with $k_0=10^{-10} \; \rm GeV$
\label{fig:3}}
\end{figure}

\newpage
\begin{figure}[t]
  \vspace*{0.0cm}
     \centerline{
         \epsfig{figure=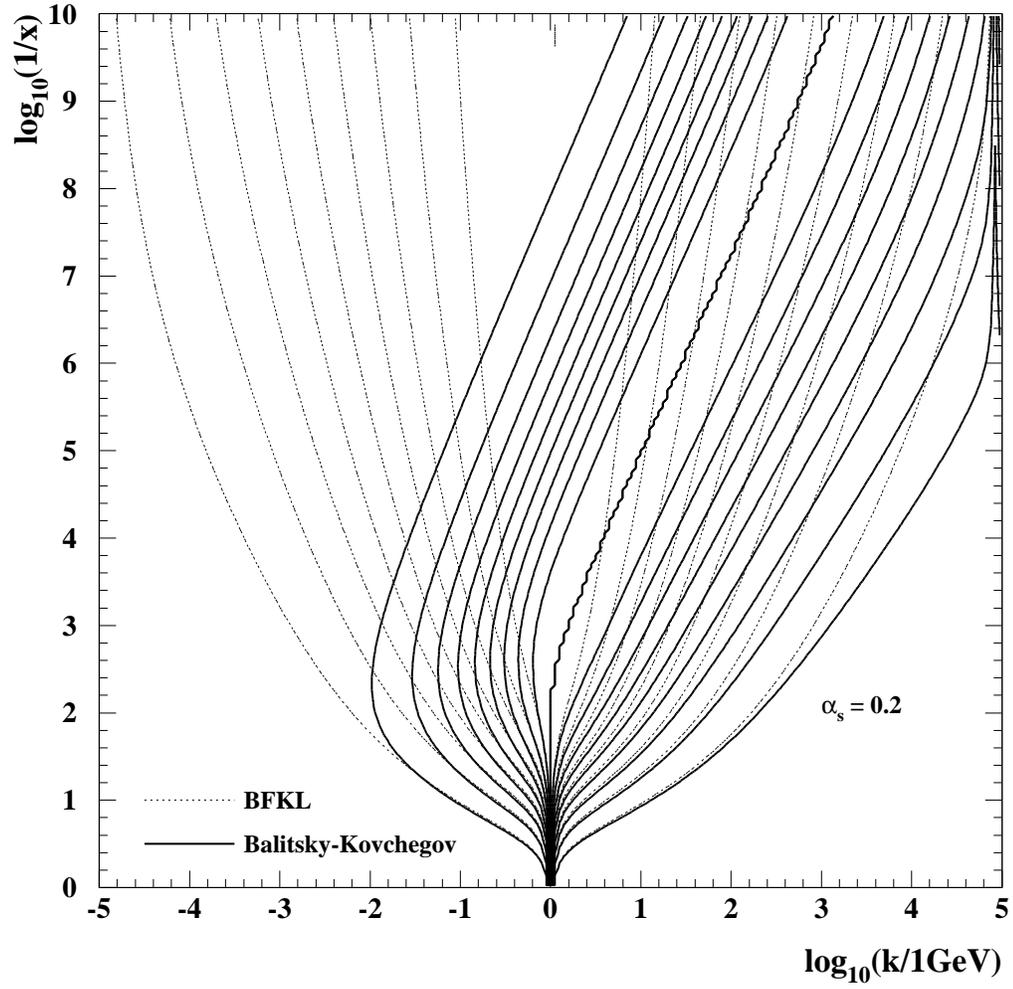,width=15cm}
           }
\vspace*{0.5cm}
\caption{\it The lines of constant values of the BFKL and the BK
re-normalized solutions  $\Psi(k,Y)$ ($Y=\ln (1/x)$) in the
$(\log_{10}(k),\log_{10}(1/x))$-plane.
\label{fig:4}}
\end{figure}

\newpage
\begin{figure}[t]
  \vspace*{0.0cm}
     \centerline{
         \epsfig{figure=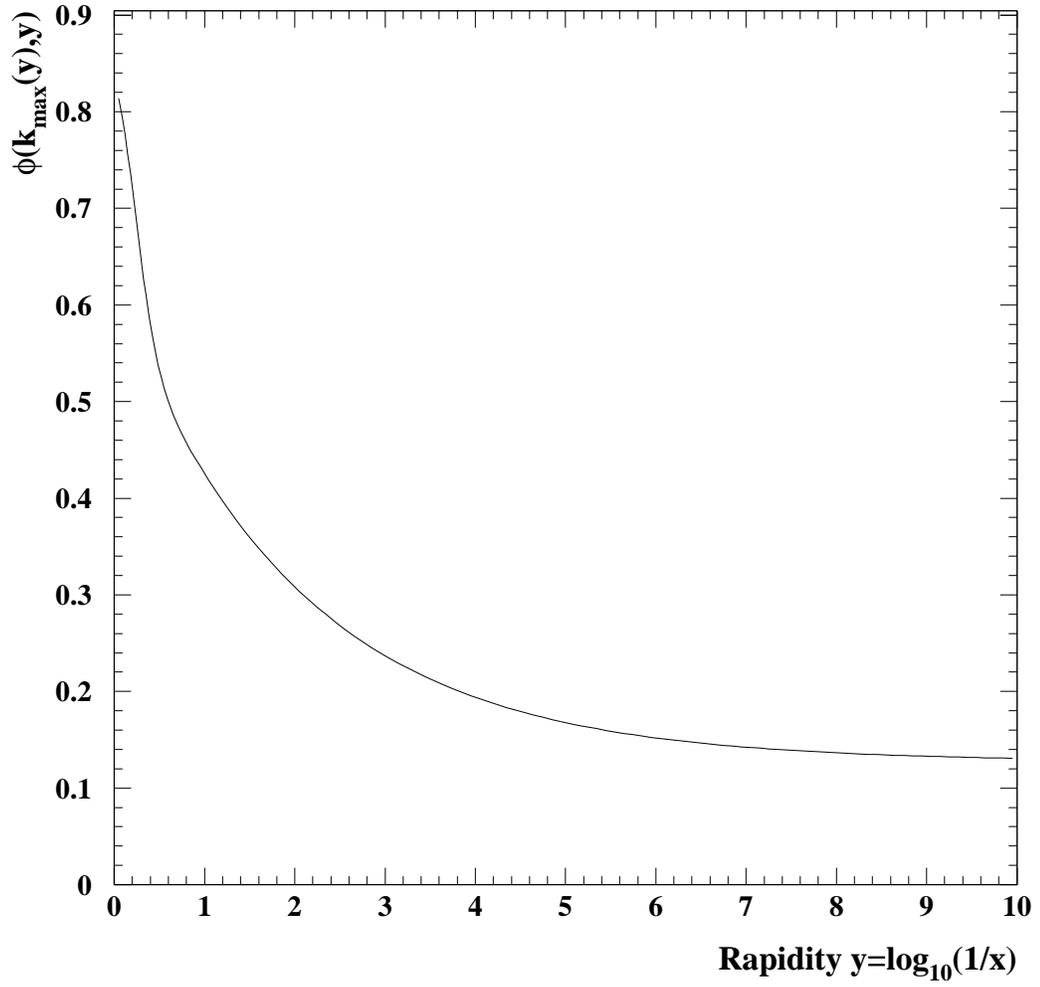,width=15cm}
           }
\vspace*{0.5cm}
\caption{\it Scaling condition (\ref{eq:crticond}) as a function of $Y$.
\label{fig:5}}
\end{figure}

\newpage
\begin{figure}[t]
  \vspace*{0.0cm}
     \centerline{
         \epsfig{figure=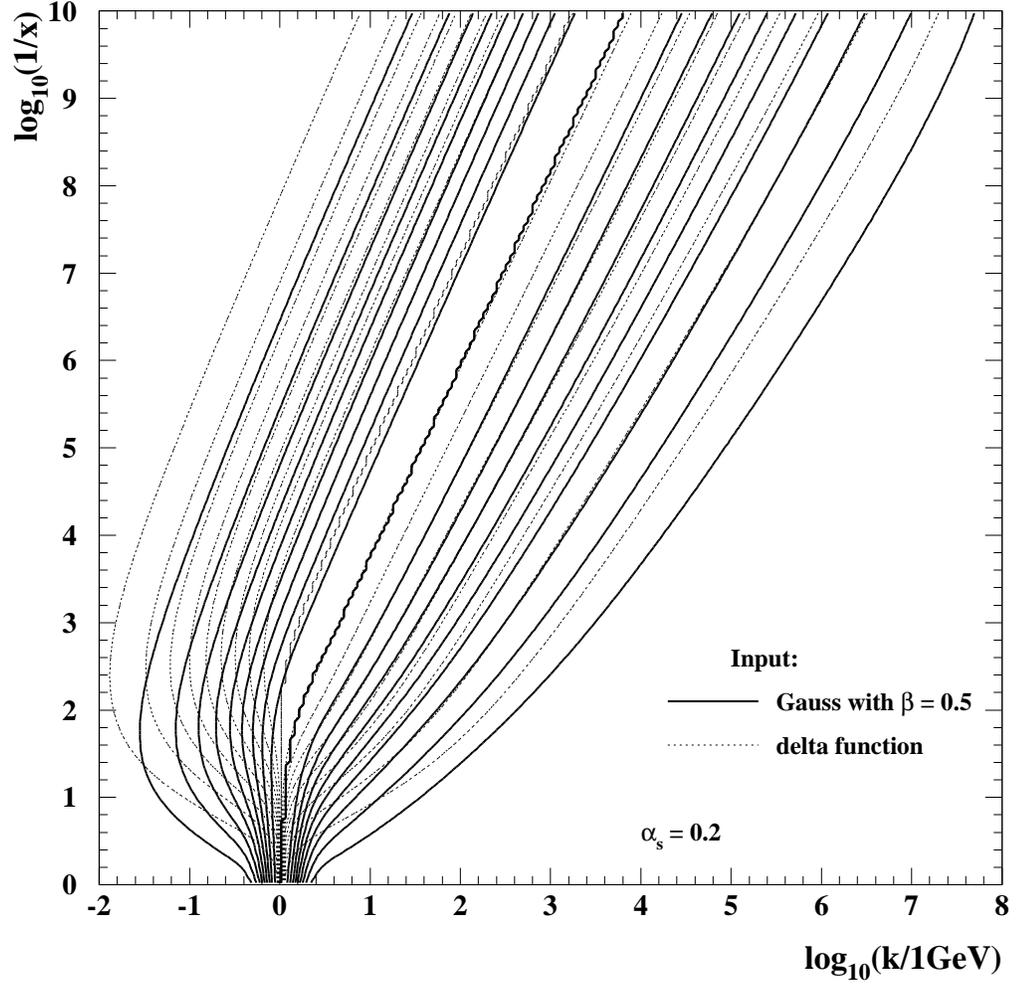,width=15cm}
           }
\vspace*{0.5cm}
\caption{\it
The lines of constant values of the Balitsky-Kovchegov re-normalized solutions
$\Psi(k,Y)$ in the $(\log_{10}(k),\log_{10}(1/x))$-plane for the delta-like input
(\ref{eq:start1})
and the Gaussian input  (\ref{eq:start2}).
\label{fig:6}}
\end{figure}

\newpage
\begin{figure}[t]
  \vspace*{0.0cm}
     \centerline{
         \epsfig{figure=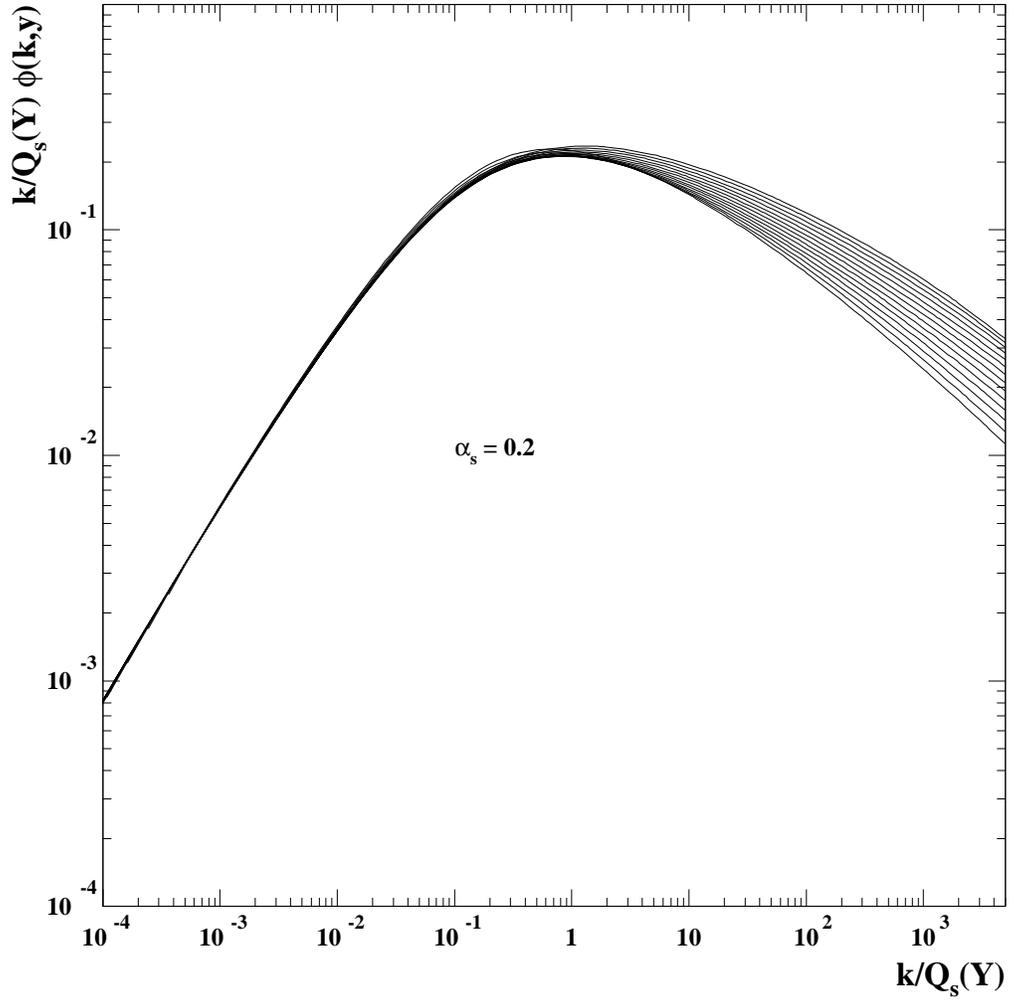,width=15cm}
           }
\vspace*{0.5cm}
\caption{
The function $(k/Q_s(Y))\,\phi(k,Y)$
plotted versus $k/Q_s(Y)$ for different values of rapidity $Y$
ranging from $10$ to $23$. The saturation scale $Q_s(Y)$ corresponds to the position
of the maximum of the function $k\,\phi(k,Y)$.
\label{fig:7}}
\end{figure}

\newpage
\begin{figure}[t]
  \vspace*{0.0cm}
     \centerline{
         \epsfig{figure=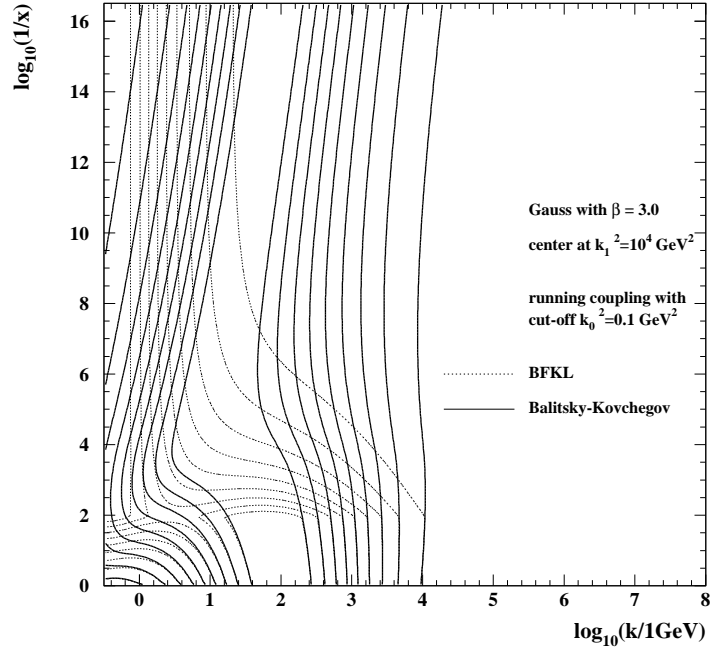,width=10cm}
           }
\vspace*{-0.5cm}
     \centerline{
         \epsfig{figure=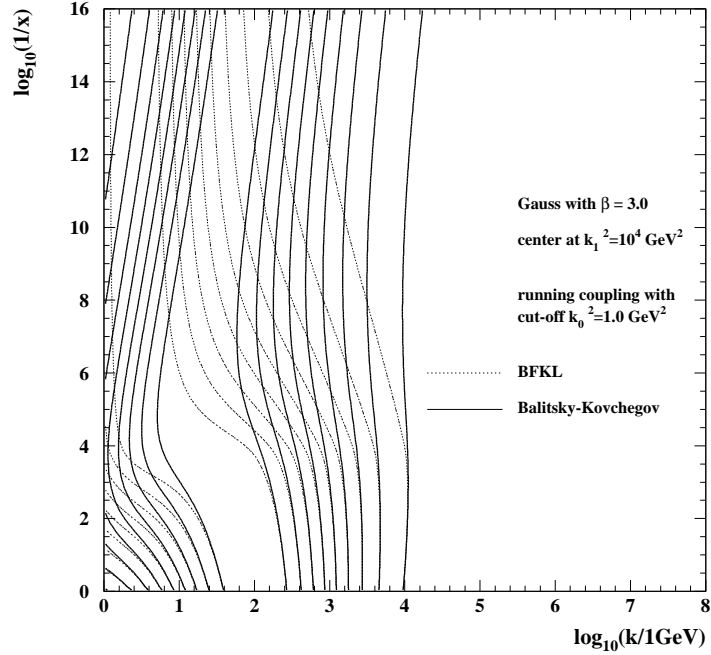,width=10cm}
           }
\vspace*{-0.5cm}
\caption{\it
The re-normalized solutions  $\Psi(k,Y)$
from the  input (\ref{eq:start1}) for
the Balitsky-Kovchegov equation with the running coupling constant
and the infra-red cut-off $k_0^2 = 0.1\, \gev^2$ (a) and $k_0^2 = 1\,\gev^2$ (b)
as functions of  $\log_{10}(1/x)$ and $\log_{10}(k/1\,\gev)$.
\label{fig:8}}
\end{figure}

 \newpage
\begin{figure}[t]
  \vspace*{0.0cm}
     \centerline{
         \epsfig{figure=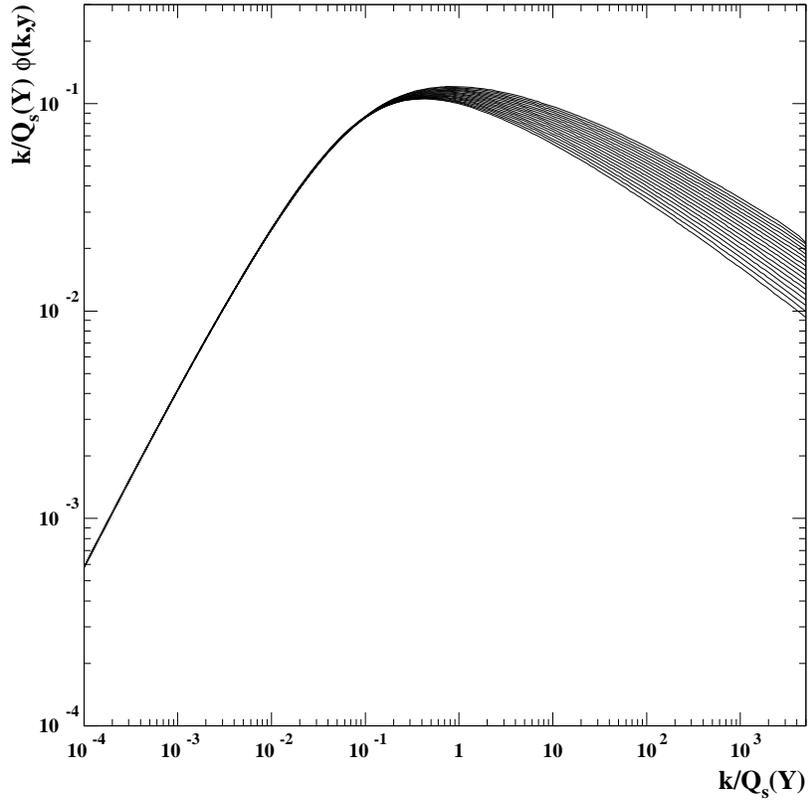,width=12cm}
           }
\vspace*{0.5cm}
\caption{\it
The function $(k/Q_s(Y))\,\phi(k,Y)$ in the  running coupling  case,
plotted versus $k/Q_s(Y)$ for different values of rapidity $Y$
ranging from $15$ to $32$. The saturation scale $Q_s(Y)$ is taken from equation
(\ref{eq:satscalerc}) with the initial condition $Q_s(Y=0)=2.0 \; GeV$.
\label{fig:9}}
\end{figure}

 \newpage
\begin{figure}[t]
  \vspace*{0.0cm}
     \centerline{
         \epsfig{figure=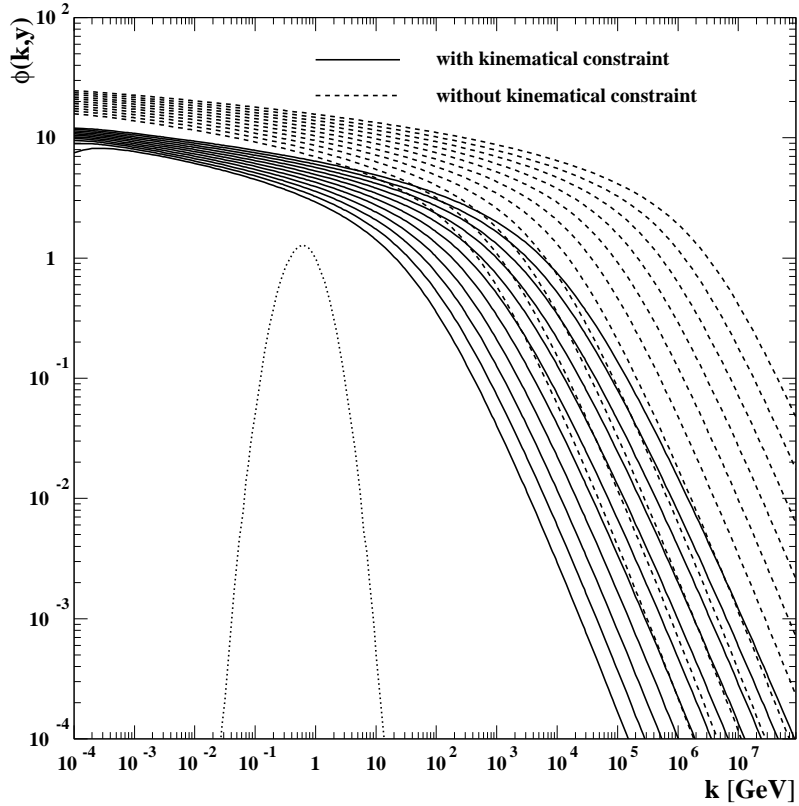,width=12cm}
           }
\vspace*{0.5cm}
\caption{\it
Comparison of solutions of the Balitsky-Kovchegov equation with and without
the kinematic constraint, plotted versus $k$ for different values of rapidity
$Y_i=20 + i*2 $, where $i$ goes from $0$ to $10$.
The  dotted line -- the  input distribution at $Y=0$; the solid line --
the result  of the calculation with the kinematic constraint, the dashed line --
the result of the calculation without the kinematic constraint.
The coupling constant $\alpha_s=0.2$ is fixed.
\label{fig:10}}
\end{figure}

\end{document}